\documentclass[prd,twocolumn, secnumarabic,preprintnumbers,amsmath,amssymb,superscriptaddress,nofootinbib,floatfix]{revtex4}
\usepackage{graphicx}
\usepackage{mathrsfs}
\usepackage{amsfonts}
\usepackage{amssymb}

\usepackage{ulem}
\usepackage[usenames]{color}
\definecolor{DarkGreen}{rgb}{0.0,0.5,0.0}

\begin{document}
\preprint{IPMU-09-0114}

\title{Observational Signatures of Gravitational Couplings in DBI Inflation}

\author{Damien A. Easson}%
\email[Email:]{easson@asu.edu}
\affiliation{Institute for the Physics and Mathematics 
of the Universe, University of Tokyo,
5-1-5 Kashiwanoha, Kashiwa, Chiba 277-8568, Japan}
\affiliation{Department of Physics, and School of Earth and Space Exploration, and Beyond Center, Arizona State University,
Tempe, AZ, 85287-1504, USA}
\author{Shinji Mukohyama}%
\email[Email:]{shinji.mukohyama@ipmu.jp}
\affiliation{Institute for the Physics and Mathematics 
of the Universe, University of Tokyo,
5-1-5 Kashiwanoha, Kashiwa, Chiba 277-8568, Japan}
\author{Brian A. Powell}
\email[Email:]{brian.powell@ipmu.jp}
\affiliation{Institute for the Physics and Mathematics 
of the Universe, University of Tokyo,
5-1-5 Kashiwanoha, Kashiwa, Chiba 277-8568, Japan}

\date{\today}
\begin{abstract}
In scalar-tensor theories the scalar fields generically couple nontrivially to gravity.
We study the observable properties of inflationary models with non-minimally coupled inflaton and Dirac-Born-Infeld (DBI) kinetic term.
Within the assumptions of the priors of our Monte-Carlo simulations we find these models can generate new interesting observable signatures. 
Our discussion focuses on string theory inspired phenomenological models of relativistic D-brane inflation.
While successful string theory constructions of ultra-violet DBI brane inflation remain elusive, we show that in suitable regions of the parameter space it is possible to use cosmological observables to probe the non-minimial coupling. 
Fortunately, the most observationally promising range of parameters include models yielding intermediate levels of non-gaussianity in the range consistent with WMAP 5-year data, and to
be constrained further by the Planck satellite.

\end{abstract}

\maketitle

\def\be{\begin{equation}}
\def\ee{\end{equation}}
\def\bea{\begin{eqnarray}}
\def\eea{\end{eqnarray}}
\def\bs{\begin{subequations}}
\def\es{\end{subequations}}
\def\half{{\frac{1}{2}}}  
\def\g{\gamma}
\def\G{\Gamma}
\def\vp{\varphi}
\def\mpl{M_{\rm Pl}}
\def\ms{M_{\rm s}}
\def\ls{\ell_{\rm s}}
\def\lp{\ell_{\rm pl}}
\def\l{\lambda}
\def\gs{g_{\rm s}}
\def\d{\partial}
\def\co{{\cal O}}
\def\sp{\;\;\;,\;\;\;}
\def\spa{\;\;\;}
\def\r{\rho}
\def\dr{\dot r}
\def\dt{\dot\varphi}
\def\e{\epsilon}
\def\k{\kappa}
\def\m{\mu}
\def\n{\nu}
\def\om{\omega}
\def\tn{\tilde \nu}
\def\p{\phi}
\def\vp{\varphi}
\def\P{\Phi}
\def\r{\rho}
\def\s{\sigma}
\def\t{\tau}
\def\x{\chi}
\def\z{\zeta}
\def\a{\alpha}
\def\b{\beta}
\def\de{\delta}

\section{Introduction}
There is substantial evidence indicating that the early Universe underwent a brief period of rapid, accelerated expansion, or inflation \cite{Guth:1980zm,Linde:1981mu,Albrecht:1982wi}. 
In addition to solving conceptual and theoretical problems of the Standard Big Bang model (such as the
monopole, flatness and horizon problems),  inflation  provides
a quantum origin for the seeds of large scale structures in the Universe and produces theoretical predictions that are supported by
detailed measurements of temperature anisotriopies in the cosmic microwave background radiation (CMB)~\cite{Komatsu:2008hk}.
But, despite its many phenomenological successes, the microphysical origin of inflation remains unclear; indeed, ever since its inception, it has been a challenge to derive successful inflationary models from fundamental particle physics~\cite{Brandenberger:2007qi}. Fortunately, early work on D-brane inflation in string theory~\cite{Dvali:1998pa,Alexander:2001ks,Dvali:2001fw,Burgess:2001fx,Brodie:2003qv,Kachru:2003sx}, has fruitfully led to promising realizations of the scenario~\footnote{For recent reviews see, e.g. \cite{McAllister:2007bg,Baumann:2009ni}.}.

Many inflationary models are driven, in part, by an inflaton field with non-canonical kinetic term, such as k-inflation \cite{ArmendarizPicon:1999rj}, Ghost inflation \cite{ArkaniHamed:2003uz}  and Dirac-Born-Infeld (DBI) inflation \cite{Silverstein:2003hf}.  In these models the speed of sound can be significantly smaller than unity. Because cosmological perturbations travel at speeds less than that of light in these models, distinct observational signals are typically generated, including 
detectable levels of non-Gaussianity, e.g. \cite{Garriga:1999vw,Silverstein:2003hf,Alishahiha:2004eh,Chen:2006nt}. As we demonstrate in this paper, non-Gaussianity is an additional observable providing new information that can be used to distinguish between certain inflationary models.

It is the nature of quantum field theory in curved spacetime that scalar fields generically couple non-minimally to gravity~\cite{Birrell:1982ix}.
In this paper we study effects of gravitational couplings in inflationary models with non-standard kinetic terms. In particular, we are interested in determining how such couplings alter cosmological observables.
Our goal is to determine if it is possible to probe such couplings using measurements of non-Gaussianity together with measurements of other standard physical quantities.
We ultimately focus on the theoretically well-motivated setting of a non-minimally coupled inflaton field in a DBI inflationary model; however, the methods we develop
are applicable to more general settings as discussed below.

An outline of the paper is as follows: In \S\ref{motive}, we provide a brief introduction to gravitational couplings in theories with non-standard kinetic terms and DBI models
of D-brane inflation. In \S\ref{analytic}, we derive analytic expectations for the observables of canonical and non-canonical models with conformal coupling.
In \S\ref{discrim}, we discuss the prospects of observationally discriminating between pure DBI theories and DBI theory with a conformal coupling term. In \S\ref{monte},
we describe the priors and general methodology used in our Monte-Carlo simulations. In \S\ref{results}, we present our interpretation and
an analytic understanding of the results of the analysis. Finally, we provide a summary of significant findings and further conclusions in \S\ref{conclude}. Technical calculations are relegated to a series of Appendices.
\section{Motivation}\label{motive}
The most general actions we would like to study are of the form
\be\label{act}
S = \int d^{4} \! x \,  \sqrt{-g} \, \Big( F(R,\vp) + P(X,\vp)  \Big)
\,,
\ee
where $X = - \frac{1}{2}  g^{\m\n} \d_\m \vp \d_\n \vp$, so that for a canonically normalized scalar field,
$P(X,\vp) = X - V(\vp)$.  $F$ is a series whose leading contribution is the Einstein-Hilbert term and further terms encompass possible gravitational 
coupling and correction terms \footnote{We work with the reduced Planck mass, written in terms of the four-dimensional Newton's constant $G_N$ as, $\mpl = 1/ \sqrt{8 \pi G_N}\simeq 2.4 \times 10^{18}$.} :

\be\label{bigf}
F(R,\vp) = \frac{R}{2} \, \vp^2 \, \left( \mpl^2  \frac{1}{\vp^2} - \xi - \xi_1 \frac{R}{\vp^2} + \cdots \right)
\,.
\ee
Gravitational couplings of the form (\ref{bigf}) are commonly generated in dynamical 4D gravity, and have
been explicitly computed for D-brane probes in curved backgrounds~\cite{Seiberg:1999xz,Fotopoulos:2002wy}, and
in D-brane inflationary models~\cite{Kachru:2003sx}. For a recent discussion of the effects of loop corrections on slow-roll inflation models 
in a power-counting formalism of effective field theory see \cite{Burgess:2009ea}.
\subsection{DBI Inflation}
From a theoretical perspective, a particularly well-motivated instance of non-standard kinetic term occurs
in D-brane inflation: the inflaton kinetic term is of the
Dirac-Born-Infeld (DBI) form~\cite{Silverstein:2003hf}:
\be\label{Pdbi}
P(X,\vp) = - f^{-1} (\vp) \sqrt{ 1- 2 f(\vp) X}
                  +  f^{-1}(\vp)  - V(\vp) 
                  \,,
 \ee
where the warp factor $f$ is related to the harmonic function of a warped compactification.
The non-standard form of the kinetic Lagrangian imposes an effective
speed limit on the inflaton field, analogous to the speed limit imposed on particle motion in 
Special Relativity. This behavior results in a new form of
slow-roll inflation even when the potential is very steep \cite{Silverstein:2003hf,Alishahiha:2004eh}.

The most well studied D-brane inflation scenario consists of a brane anti-brane pair embedded in
the warped throat region of a conformally Calabi-Yau flux compactification in Type IIB string theory. 
The warping can be used to obtain a large hierarchy in scales between the electroweak and Planck scales. If the throat is the warped deformed conifold~\cite{KSgeom}, it is well approximated by simple Anti-de-Sitter (AdS) space
far from the tip. In the case of ultra-violet (UV) brane inflation, a mobile D3 brane falls from the UV end of the warped throat 
towards the infra-red (IR) end of the throat where an anti-D3 brane is located. The position of the D3
brane in the extra dimensions acts as an inflaton field as the brane moves toward the anti-brane
under the influence of an attractive Coulomb force. Inflation ends
via a tachyonic instability when the brane crashes into the anti-brane. The 
formal construction is now well known and we do not review the details of the model here.  The curious
reader is referred to the recent reviews \cite{McAllister:2007bg,Baumann:2009ni}. The most successful
versions of this scenario are slow-roll inflationary models. We briefly comment
on the difficulties involved in building rigorous models of UV DBI inflation within the above setting.

In the state-of-the-art constructions of D-brane inflation the inflaton field range in Planck units is 
bounded \cite{Chen:2006hs,Baumann:2006cd} by
\be\label{bmbound}
\frac{\Delta \vp}{\mpl} \leq \frac{2}{\sqrt{N_0}}
\,,
\ee
where $N_0 \gg 1$ is an integer representing the flux stabilizing the throat. The relatively few models capable of 
satisfying this bound and producing observable levels of non-Gaussianity typically predict a blue spectral index and
a negligible tensor component \cite{Peiris:2007gz}.  
Finally, DBI constructions must overcome difficult backreaction constraints~\cite{Silverstein:2003hf,Easson:2007dh,Bean:2007eh,Leblond:2008gg}.

In this paper we are interested in the general observational properties catalyzed by combining the DBI action with gravitational couplings,
and work within a phenomenological setting inspired by the afore mentioned brane inflation scenario. It is our modest goal to
ascertain the possibility of observing gravitational couplings in inflationary models with non-standard kinetic terms and
not to build realistic particle physics models. While all of the presented solutions conform to the latest observational constraints imposed by
the WMAP satellite~\cite{Komatsu:2008hk}, the majority do not obey the constraint  on the field range (\ref{bmbound}). Interesting models of 
inflation in string theory capable of avoiding this bound have recently been proposed in \cite{Becker:2007ui,Silverstein:2008sg,McAllister:2008hb,Avgoustidis:2008zu}. From the effective field theory standpoint,
the DBI action is of particular interest even divorced from a particular string theory model as it represents the minimal realization of non-linear Lorentz invariance.
\subsection{Equations of Motion and Gravitational Coupling}
Variation of the action (\ref{act}) using (\ref{Pdbi}) with  respect to the metric $g^{\m\n}$ leads to the generalized Einstein equation:
\begin{eqnarray}
F_R \, R_{\m\n} - \half g_{\m\n} F + g_{\m\n} \Box F_R - \nabla_\m \nabla_\n F_R  =  \nonumber \\
  \frac{\d_\m \vp \, \d_\n \vp -  g_{\m\n} \,\left(\d^\s\vp \,\d_\s \vp 
+ f^{-1}(\vp)\right)}{\sqrt{1 + f (\vp)g^{\m\n} \d_\m \vp \, \d_\n \vp}}  \nonumber \\
+ \left(f^{-1}(\vp) - V(\vp) \right) g_{\m\n} 
\,,
\end{eqnarray}
where $F_R \equiv \d F/\d R$. The equation of motion for the field $\vp$ is:
\begin{eqnarray}
\nabla_\m ( \gamma \, \d^\m \vp )  + f^{-2} f' (\gamma^{-1} - 1) \nonumber \\
 -  \frac{1}{2} f^{-1} f'  \gamma g^{\m\n}\partial_{\m}\vp \, \partial_{\n} \vp & - & V' - F' =0
\,,
\end{eqnarray}
where $\prime \equiv \d/\d\vp$,
and we have defined
\be\label{gamma}
\gamma \equiv \frac{1}{\sqrt{1 + f(\vp) g^{\m\n}\, \partial_{\m}\vp \, \partial_{\n} \vp }}
\,.
\ee
In this paper we focus on a particular well-motivated and simple form for the gravitational coupling 
and truncate $F(R,\vp)$ at the second term in (\ref{bigf}) . This term, $-\frac{1}{2} \xi R \vp^2$, plays a significant role in brane 
inflation since the field associated to the brane position is a conformally coupled scalar~\cite{Seiberg:1999xz, Silverstein:2003hf,Kachru:2003sx}.
The term is ubiquitous in effective field theories of quantum fields in curved spacetime~\cite{Birrell:1982ix}, and is renormalizable by power
counting arguments.
\section{Analytic Analysis}\label{analytic}
\subsection{Canonical Models}
The effects of non-minimal couplings of the form $\xi R \varphi^2$ were
first investigated in the context of canonical inflation, described by the action,
\begin{eqnarray}
\label{jordan}
S &=& \int d^4x \sqrt{-g}\left\{\frac{M_{\rm Pl}^2}{2}R - \frac{\xi}{2}R\varphi^2 \right. \nonumber \\
&&- \left. \frac{1}{2}g^{\mu\nu}\partial_\mu \varphi \partial_\nu \varphi - V(\varphi) \right\} \,,
\end{eqnarray}
in which the kinetic term of the field is canonically normalized,
$\dot{\varphi}^2/2$.  
From (\ref{jordan}) we see that introducing the non-minimal coupling term generates an effective Planck mass of the form
\be
M_{eff} \equiv \sqrt{\mpl^2 - \xi \vp^2}
\,.
\ee
For field values larger than the critical value,
\be\label{phic}
\vp_c = \frac{\mpl}{\sqrt{\xi}}
\ee
the effective Newton's constant can become negative.~\footnote{For the present discussion we will always take $\xi \geq 0$. Subsequently, we will focus on the conformal value $\xi =1/6$ (cf. discussion at the end of  \S\ref{discrim}).} In a cosmological
background with non-minimally coupled scalar (and positive $\xi$) the anisotropic shear diverges without bound as $\vp$ approaches $\vp_c$~\cite{Starobinsky81, Futamase:1989hb}. We will regard
such large field values as outside the range of validity of our effective theory.  In all of the following analysis we ensure that $\vp < \vp_c$.

The observables of such theories are typically computed in the conformally-related
Einstein frame, in which the inflaton is minimally coupled to gravity.  
The Einstein frame action is obtained by performing the well known Weyl rescaling
of the metric,
\begin{equation}\label{ct}
\tilde{g}_{\mu \nu} = \Omega^2 g_{\mu \nu},
\end{equation}
where $\Omega^2 =1-\xi\frac{\varphi^2}{M_{\rm Pl}^2}$, followed by a field redefinition, $\varphi \rightarrow \sigma$  (see Appendix~\ref{ctrans}) so that
the action becomes

\begin{equation}
\label{ef}
S = \int d^4x \sqrt{-\tilde{g}}\left\{\frac{M_{\rm Pl}^2}{2}\tilde{R} - \frac{1}{2}\tilde{g}^{\mu
\nu}\partial_\mu \sigma \partial_\nu \sigma - \tilde{V}(\sigma)\right\},
\end{equation}
where $\tilde{V}(\sigma) = V(\varphi)\Omega^{-4}(\varphi)$.
While this conformally related theory is physically distinct from the original Eq. (\ref{jordan}), the cosmological observables
calculated in this frame are equivalent \cite{Makino:1991sg}.  Therefore, we conclude that \it any
non-minimally coupled canonical single-field theory is observationally equivalent to some minimally coupled canonical
theory \rm (within the range of validity of the transformation (\ref{ct})). This makes it impossible to observationally determine the nature of the gravitational inflaton coupling in canonical theories.  The result is in agreement with the analysis of \cite{Creminelli:2003iq}. 
\subsection{Non-Canonical Models}
We next consider the case of a non-canonical inflaton.  In what follows,
we study
DBI inflation \cite{Silverstein:2003hf,Alishahiha:2004eh} as a prototype non-canonical theory.  The standard minimally coupled DBI action is 
\begin{eqnarray}
\label{mindbi}
S &=& \int d^4x \sqrt{-g}\left\{\frac{M_{\rm Pl}^2}{2}R  - \right.\\ \nonumber
 && \left.\frac{1}{f(\theta)}\left[\sqrt{1+f(\theta)g^{\mu \nu}\partial_\mu \theta \partial_\nu \theta} -1\right]
- V(\theta)\right\}.
\end{eqnarray}
In brane inflation $f(\theta)$ is the warp factor of the compactified geometry and $V(\theta)$ is the potential that arises from
the presence of anti-D3-branes and other perturbative and non-perturbative sources.
The action for single field DBI inflation with non-minimal coupling is 
\begin{eqnarray}
\label{dbi}
S &=& \int d^4x \sqrt{-g}\left\{\frac{M_{\rm Pl}^2}{2}R - \frac{1}{2}\xi R\phi^2 - \right.\\ \nonumber
 && \left.\frac{1}{f(\phi)}\left[\sqrt{1+f(\phi)g^{\mu \nu}\partial_\mu \phi \partial_\nu \phi} -1\right]
- V(\phi)\right\}.
\end{eqnarray}
As in the case of the canonical theory, we can remove the explicit nonminimal coupling via the transformation
(\ref{ct}).  This results in the action 
\be\label{actp}
S = \int d^{4} \! x \,  \sqrt{- \tilde g} \, \left(\frac{ \mpl^2}{2} \tilde R + P(  X,\phi) \right)
\,,
\ee
where 
\be
X = - \frac{1}{2} \tilde g^{\m\n} \d_\m\phi \d_\n \phi
\,.
\ee
The corresponding functional form for $P$ is:
\bea\label{xip}
\left(1-\xi \frac{\phi^2}{\mpl^2} \right)^2 \, P(\phi,  X) = 6 \xi^2   X \frac{\phi^2}{\mpl^2} && \\ \nonumber
- f^{-1}(\phi) \sqrt{1 - 2 X f \left(1-\xi \frac{\phi^2}{\mpl^2} \right)} &+& f^{-1}(\phi) - V(\phi)
\,.
\eea
Therefore, in terms of a homogeneous field $\phi(t)$, the Lagrangian (\ref{actp}) becomes:
\begin{eqnarray}
\label{pdbi}
\frac{\mathcal{L}}{ \sqrt{- \tilde g} }= \frac{ \mpl^2}{2} \tilde R &+& g(\phi,\dot{\phi}) + \\
\Omega^{-4}\Big[\Big(1& - & \sqrt{1-\dot{\phi}^2f(\phi)\Omega^2} \Big)\frac{1}{f(\phi)} - V(\phi)\Big]
\nonumber 
\end{eqnarray}
where
\begin{equation}
\label{g}
g(\phi,\dot{\phi}) = 3\xi^2\frac{\dot{\phi}^2}{\Omega^4}\frac{\phi^2}{M_{\rm Pl}^2}.
\end{equation}
The presence of the function $ g(\phi,\dot{\phi})$ makes it impossible to fully duplicate the observables derived from 
(\ref{dbi}) using observables derived from a standard minimally coupled DBI theory (\ref{mindbi}). 
Because minimally coupled DBI inflation makes specific observational predictions
\cite{Bean:2007hc,Peiris:2007gz}, it is possible that this
departure from a pure DBI theory might lead to distinct observable physics.  As a first step towards exploring this possibility, we examine  
Eq. (\ref{pdbi}) under various field redefinitions.
\subsubsection{Non-Relativistic Limit}
We first redefine the scalar field so that its kinetic term reduces to the canonical form in the
non-relativistic limit, $f(\phi)\dot{\phi}^2 \ll 1$.   
In the non-relativistic limit the
action (\ref{pdbi}) is reduced to 
%
\begin{eqnarray}
\label{dbinr}
 S_{nr} &=& \int d^4x \sqrt{-\tilde{g}} \left[\frac{M_{\rm Pl}^2}{2}\tilde{R} \right. \nonumber \\
  &-&\left.\frac{1-\xi(1-6\xi)\frac{\phi^2}{M_{\rm Pl}^2}}{2\left(1-\xi\frac{\phi^2}{M_{\rm Pl}^2}\right)^2} \tilde{g}^{\mu\nu}\partial_{\mu}\phi\partial_{\nu}\phi - \frac{V}{\Omega^4} \right].
\end{eqnarray}
Thus, we define a new field $\Phi_c$ by 
%
\begin{equation}
\label{Phi_c}
 d\Phi_c = 
   \frac{\sqrt{1-\xi(1-6\xi)\frac{\phi^2}{M_{\rm Pl}^2}}}
   {1-\xi\frac{\phi^2}{M_{\rm Pl}^2}}d\phi,
\end{equation}
so that 
%
\begin{equation}\label{nrform}
 S_{nr} = \int d^4x \sqrt{-\tilde{g}}
  \left[\frac{M_{\rm Pl}^2}{2}\tilde{R}
   -\frac{1}{2}
   \tilde{g}^{\mu\nu}\partial_{\mu}\Phi_c\partial_{\nu}\Phi_c
   - \tilde{V}(\Phi_c) \right],
\end{equation}
where $\tilde{V}(\Phi_c)\equiv V(\phi)/\Omega^4(\phi)$. 
Taking $\xi=1/6$, we can integrate (\ref{Phi_c}) to obtain the relation,
\begin{equation}
 \frac{\phi}{M_{\rm Pl}} =
\sqrt{6} \tanh \left( \frac{\Phi_c}{\sqrt{6} M_{\rm Pl} } \right). 
\end{equation}
Since the minimally coupled DBI theory (\ref{mindbi}) reduces to the same non-relativistic form as (\ref{nrform}), we expect no physical differences between the theories in this limit.
\subsubsection{Ultra-Relativistic Limit}
We next consider a different field redefinition which brings Eq. (\ref{pdbi}) into physical agreement with
minimally coupled DBI inflation when the field is highly relativistic.  In this
regime, where the expression inside the square root becomes small,  the warp factor plays a dominant role in determining the dynamics.  This suggests 
a field redefinition $\phi \rightarrow \Phi$ that brings the expression inside the square root
into the form present in  minimally coupled DBI, $1 - \dot{\Phi}^2f(\Phi)$.
Thus, we define a new field, 
\begin{equation}
 \frac{\sqrt{\xi}\phi}{M_{\rm Pl}} =
  \sin\left(\frac{\sqrt{\xi}\Phi}{M_{\rm Pl}}\right)
\end{equation}
so that
%
\begin{equation}
 \tilde{\cal L} = \tilde{\cal L}_{DBI} + g(\Phi,X),
  \label{eqn:action-good-for-DBI-regime}
\end{equation}
where 
%
\be
 \tilde{\cal L}_{DBI}  = 
   \frac{1}{\tilde{f}}
   \left[ 1- \sqrt{1-2\tilde{f}X} \right] - \tilde{V}(\Phi_c), 
   \label{LDBI}
   \ee
   \be
g(\Phi,X)  =  6\xi X
  \tan^2\left(\frac{\sqrt{\xi}\Phi}{M_{\rm Pl}}\right),
\ee
and 
%
\begin{equation}
  X  =
  -\frac{1}{2}\tilde{g}^{\mu\nu}
  \partial_{\mu}\Phi\partial_{\nu}\Phi.
\end{equation}
Here, $\Phi_c$ in the potential $\tilde{V}(\Phi_c)$ must be expressed in
terms of $\Phi$ by eliminating $\varphi$ from definitions of $\Phi$ and
$\Phi_c$, i.e. $\tilde{V}(\Phi_c(\Phi))$. For $\xi=1/6$ we have the
relation 
%
\begin{equation}
\tanh\left(\frac{\Phi_c}{\sqrt{6}M_{\rm Pl}}\right) 
= \sin\left(\frac{\Phi}{\sqrt{6}M_{\rm Pl}}\right). 
\end{equation}
In the relativistic limit, $2\tilde f X \approx 1$, and since $6\xi \tan^2(\sqrt{\xi}\Phi/M_{\rm Pl}) \ll 1$ for $\xi
\sim \mathcal{O}(10^{-1})$ and $\Phi \sim \mathcal{O}(M_{\rm Pl})$ (parameter magnitudes of interest in this
paper), $g$ is small and Eq. (\ref{eqn:action-good-for-DBI-regime}) reduces to the minimally coupled DBI Lagrangian
(\ref{mindbi}). We elaborate on this issue in \S~\ref{results}.
\subsubsection{Intermediate Regime}
Since the above two field redefinitions result in a conformal theory that is almost of the same form as a minimally coupled
DBI theory in the non-relativistic and relativistic regimes, respectively, we expect both theories to be
observationally degenerate in these limits.  However, it is unclear what the nature of the theory given
by Eq. (\ref{pdbi}) is in the
intermediate regime.  Might there be a region of parameter space for which this theory gives different
predictions from a minimally coupled DBI theory?  If DBI is the presumed inflationary theory, any
observational differences between the two theories could yield direct information about the presence of
gravitational couplings, in contrast to the case of canonical models.  In the next section, we pursue this question in detail. 
\section{Observationally discriminating nonminimal theories}\label{discrim}
\subsection{Non-Gaussianity}
Non-canonical inflation models have the defining characteristic that
inflaton fluctuations generally propagate at speeds less than that of
light.  As a result, the fluctuations are non-Gaussian, at levels significantly larger than
the negligible levels produced in single field slow-roll inflation \cite{Maldacena:2002vr}.
One non-Gaussian statistic is the three-point function of the co-moving curvature perturbation, $\zeta$, or its fourier transform, the bispectrum~\cite{Gangui:1993tt,Komatsu:2001rj}:
\be
\left<\zeta({{\bf k}_1})\zeta({{\bf k}_2})\zeta({{\bf k}_3})\right> 
\,.
\ee
The bispectrum is characterized by its
amplitude, shape, and scale dependence as a function of comoving wavenumbers $k_1$, $k_2$, and $k_3$. For the leading-order contributions to the non-Gaussianities in general
non-canonical theories we define the estimators \cite{Chen:2006nt},~\footnote{Our sign conventions for $f_{NL}$ are opposite to those of the WMAP team.} 
\begin{eqnarray}
\label{fl}
f^{\lambda}_{NL} &=& - \frac{5}{81}\left(\frac{1}{c^2_s} - 1 - 2\Lambda\right),\\
\label{fc}
f^{c}_{NL} &=&  \frac{35}{108}\left(\frac{1}{c_s^2} -1\right),
\end{eqnarray}
where
\begin{equation}
\Lambda = \frac{XP_{,XX} + \frac{2}{3}X^2P_{,XXX}}{P_{,X} + 2XP_{,XX}},
\end{equation}
with $P$ the fluid pressure.  The hydrodynamical sound speed is defined
\begin{equation}
\label{sound}
c^2_s = \frac{P_{,X}}{P_{,X} + 2XP_{,XX}}. 
\end{equation}
These two contributions correspond to different shapes in Fourier space \cite{Babich:2004gb,Chen:2006nt}.  In
the case of DBI inflation, $f^{\lambda}_{NL}$ is identically zero, and so an experimental detection of the shape
of the non-Gaussianities offers the
possibility of distinguishing DBI inflation from more general non-canonical scenarios, in particular, its nonminimally
coupled counterpart.  
However, both shapes are maximized for equilateral configurations ($k_1=k_2=k_3$) and will be
difficult to distinguish in practice unless there is a detection of high significance \cite{Fergusson:2008ra}.
We will therefore not consider using the shape of non-Gaussianities as a discriminator.
Nevertheless, in \cite{Easson:2009kk}, it was suggested that since the speed of sound in the non-minimal DBI models differs from that
of pure DBI, it may be possible to distinguish between the two theories through cosmological observables.
\subsection{Observables}
We examine the full set of observables for each model, including the
amplitude of non-Gaussianities as well as power spectra.
We model the power spectrum as a power-law + running,
\begin{equation}
P_{\mathcal R}(k) = A_s \left(\frac{k}{k_0}\right)^{1 - n_s + \frac{1}{2}\alpha {\rm
ln}\left(\frac{k}{k_0}\right)},
\end{equation}
and include a tensor contribution.  
We adopt a Monte Carlo approach \cite{Easther:2002rw,Peiris:2007gz,Powell:2007gu} -- we stochastically generate large numbers of
inflation models of both types (minimal and nonminimal) and calculate their
observables, $(n_s,\, \alpha,\, r,\,f_{NL})$.
We then compare the distribution of predictions of each theory in the observable parameter space, and identify regions that are populated by one model and not the
other.   One can then discuss the plausibility of distinguishing
these models with future experiments.

General non-canonical inflation leads to a power spectrum of primordial curvature perturbations of
amplitude \cite{Garriga:1999vw},
\begin{equation}
\label{dens}
P_{\mathcal R}(k) = \frac{1}{8\pi^2 M_{\rm Pl}^2}\frac{H^2}{c_s \epsilon},
\end{equation}
and a spectrum of gravitational waves of amplitude,
\begin{equation}
\label{tens}
P_h(k) = \frac{2}{\pi^2}\frac{H^2}{M_{\rm Pl}^2}.
\end{equation}
Here, $c_s$ is as defined in Eq. (\ref{sound}), and 
\begin{equation}
\epsilon = \frac{XP_{,X}}{M_{\rm Pl}^2 H^2}.
\end{equation}
Because the density perturbations travel at speed $c_s$, Eq. (\ref{dens}) is to be evaluated at
sound horizon crossing, $c_s k = aH$, for each comoving wavenumber $k$. 
However, because the tensor perturbations propagate at the speed of light,  Eq. (\ref{tens}) is
evaluated when $k = aH$, with the result that scalar and tensor perturbations on the same comoving
scale $k$ exit the horizon at different times.  
Taking this difference into account can be important \cite{Lorenz:2008je, Lorenz:2008et, Agarwal:2008ah, Powell:2008bi}.
\begin{figure*}[htp]
\centering
$\begin{array}{cc}
\includegraphics[width=0.49 \textwidth,clip]{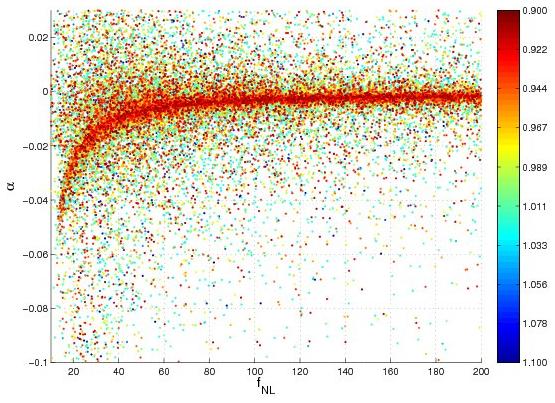}
\includegraphics[width=0.49 \textwidth,clip]{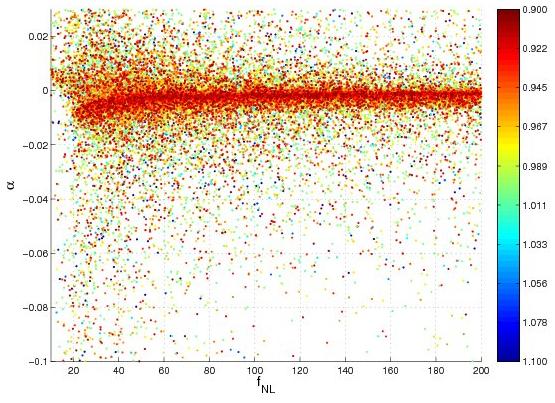}
\end{array}$
\caption{Monte Carlo results for 60,000 models each of DBI with conformal coupling (left) and
minimal coupling (right).  Coloring denotes the value of $n_s$. Observables are quoted at $k =
0.002\,h{\rm Mpc}^{-1}$.}
\end{figure*}
For a given scale $k$, the tensor-to-scalar ratio 
\be
r \equiv \frac{P_h}{P_\mathcal{R}}
\,,
\ee
 can be evaluated at any time between this mode's scalar
crossing and tensor crossing, since the quantities in it are assumed constant over this time
period. 
For the solutions studied in this analysis, we have verified that the difference in horizon crossings is not important ($H_t \approx H_s$).  In this case we may use the familiar 
\be\label{eqn:r_cs_epsilon}
r = 16\epsilon c_s.
\ee
Additionally, we define the scalar spectral index, 
\begin{equation}
\label{spindex}
n_s - 1 = \frac{d {\rm ln}P_{\mathcal{R}}(k)}{d{\rm ln}k} = -2\epsilon -\eta -s,
\end{equation}
where 
\begin{eqnarray}
\label{eta}
\eta &=& \frac{\dot{\epsilon}}{\epsilon H},\\
\label{s}
s &=& \frac{\dot{c_s}}{c_s H}.
\end{eqnarray}
The running of the spectral index, $\alpha = \frac{dn_s}{d{\rm ln}k}$, can be obtained from the
above via the relation,
\begin{equation}
\frac{d{\rm ln}k}{d\phi} = M_{\rm Pl}\frac{\sqrt{2\epsilon}}{\epsilon - 1}.
\end{equation}
In the conventions we are using, the action (\ref{jordan}) (without potential), 
is conformally invariant for the special value $\xi = \frac{1}{6}$. The presence of a general
potential or a non-standard kinetic term (such as in DBI) will break this symmetry.
When it is necessary for us to consider an actual numerical value for the non-minimal
coupling parameter, we shall take it to be $\xi = \frac{1}{6}$. We may also casually refer to
the case of non-minimal coupling as the \it conformal \rm case.
\section{Monte-Carlo Analysis}\label{monte}
\subsection{Equations of Motion}
We obtain equations of motion for the minimal models from the Lagrangian (\ref{mindbi}),
\begin{eqnarray}
\label{eomm}
3M_{\rm Pl}^2H^2 &=&  \rho \nonumber\\
&=& f^{-1}(\gamma-1)+V,\nonumber\\
2M_{\rm Pl}^2\dot{H} &=& -(\rho + P) \nonumber \\
&=& \gamma f^{-1}({\gamma}^{-2} -1).
\end{eqnarray}
The equations of motion for the conformal models are obtained from Eq. (\ref{eqn:action-good-for-DBI-regime}),
\begin{eqnarray}
\label{eom}
3M_{\rm Pl}^2 \tilde H^2 &=&  \tilde \rho \nonumber\\
&=& \tilde f^{-1}(\tilde \gamma-1)+g_{,X} X + \tilde V,\nonumber\\
2M_{\rm Pl}^2\dot{\tilde H} &=& - (\tilde \rho + \tilde P) \nonumber \\
&=& -  \left( \gamma + g \right)X.
\end{eqnarray}
where $\tilde{\gamma} = (1-2\tilde{f}X)^{-1/2}$.  
\subsection{Priors on $V$ and $f$}
Care should be taken to appropriately determine the priors on the functions $V(\theta)$, $\tilde{V}(\Phi)$, and $f(\theta)$,
$\tilde{f}(\Phi)$.  We  must ensure that any observable differences found between
these models is truly a result of physics and not a poor choice of priors for these functions.  Note
that Eqs. (\ref{eomm}) are equivalent to Eqs. (\ref{eom}) when $g = 0$, $V(\theta) \simeq
\tilde{V}(\Phi)$, and $f(\theta) \simeq \tilde{f}(\Phi)$. This is because the Lagrangian Eq. (\ref{LDBI})
coincides with that of minimally coupled DBI in the relativistic limit, when the warp factor, $f$, is important.
Therefore, to minimize physical differences, we should choose the same priors for $f(\theta)$ and
$\tilde{f}(\Phi)$.  However, Eq.
(\ref{LDBI}) is not equivalent to the minimally coupled Lagrangian in the non-relativistic limit, when $V$ is
important.  Therefore, we cannot impose the prior $V(\theta) \leftrightarrow \tilde{V}(\Phi)$.  Rather, we must
appeal to the Lagrangian (\ref{dbinr}), which {\it does} coincide with minimally coupled DBI in the
non-relativistic model.  This suggests the prior $V(\theta) \leftrightarrow \tilde{V}(\Phi_c)$.  
%
\subsection{Parameter Values and Methodology}
With the proper priors in place, we now define the functional forms of the potentials and the warp factors.   
For the potentials we consider the Taylor expansion,
\begin{equation}
\label{pot}
V(\theta) = \sum_{n=0}^{\infty}\frac{V_{n}}{2n!}\left(\frac{\theta}{M_{\rm Pl}}\right)^n,
\end{equation}
\begin{equation}
\tilde V(\Phi_c) = \sum_{n=0}^{\infty}\frac{\tilde V_{n}}{2n!}\left(\frac{\Phi_c}{M_{\rm Pl}}\right)^n,
\end{equation}
truncated at $n=4$.  Fourth order is about the extent to which we ever expect to reliably reconstruct the inflaton potential, and hence this form is motivated by phenomenology. We assume that the warp factors are of the AdS form, $f(\theta) = \lambda/\theta^4$,
$\tilde f(\Phi) = \tilde \lambda/\Phi^4$.
We draw the potential parameters from the ranges $|V_n/V_0|$, $|\tilde{V}_n/\tilde{V}_0| \in [10^{-5},10^5]$, where
$V_0,\tilde{V}_0$ is determined by normalizing the spectra to the best-fit WMAP5 value, $A_s = 3.2\times
10^{-9}$ \cite{Komatsu:2008hk}.  The warp factor is determined by drawing the rescaled warping from the range
$\lambda/V_0$,
$\tilde{\lambda}/\tilde{V}_0 \in [10^{-5},10^3]$.
We consider random initial conditions for $\dot{\theta}_i$ (and
$\dot{\Phi}_i$)
as well as the initial field values, $\theta_i$ (and $\Phi_i$).  While the minimally coupled field
$\theta$ is unbounded, the conformally coupled field must satisfy $\Phi_i <
\sqrt{\frac{3}{2}}\pi M_{\rm Pl}$, corresponding to the bound (\ref{phic}).  
The Monte Carlo analysis is carried out for the minimal models by numerically solving Eqs (\ref{eomm}), and
for the conformal models by solving Eqs. (\ref{eom}). 

Each solution is evolved
forward in time until sufficient inflation is obtained, here chosen to be $N = 55$ e--foldings.  The
number of e--foldings is obtained by solving the equation
\begin{equation}
\label{efolds}
dN = - H dt
\end{equation}
along with each system. Models which yield insufficient inflation are rejected, as
are those that evolve to a meta-stable minimum and eternally inflate.  For solutions that reach $N
= 55$, by cutting off the evolution at this point we are supposing that inflation ends via 
tachyonic instability in some auxiliary field.   

The correspondence between the comoving scale, $k$, and the number of efolds of inflation, $N$, is
different for the scalar and tensor modes because of the difference in horizon crossing times.  
It is determined by solving
\begin{equation}
\frac{d(aH)}{dt} = (1-\epsilon)aH^2,
\end{equation}
with $k = aH$ for tensors and $k c_s = aH$ for scalars.
However, as mentioned, the difference in crossing times can be neglected, and so we evaluate all observables at $aH = 0.002 h{\rm Mpc}^{-1}$.
\begin{figure*}[htp]
\centering
$\begin{array}{cc}
\includegraphics[width=0.49 \textwidth,clip]{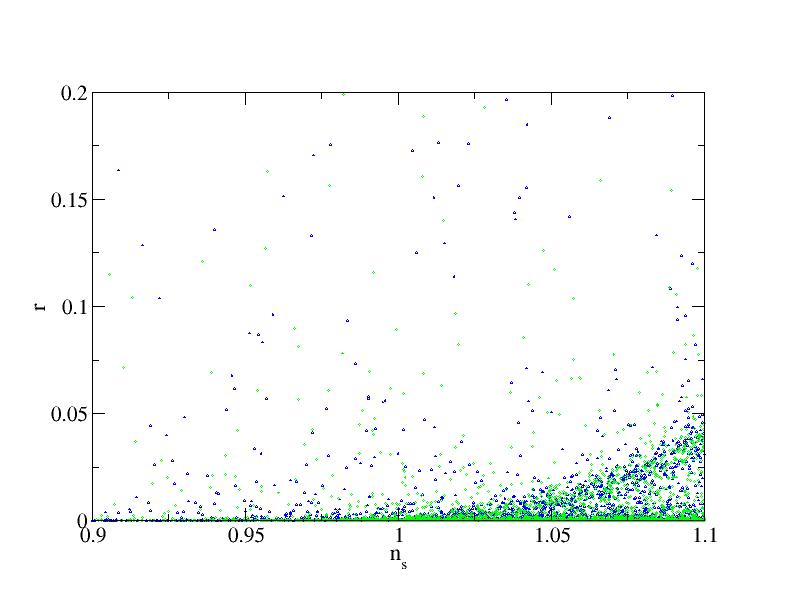}
\includegraphics[width=0.49 \textwidth,clip]{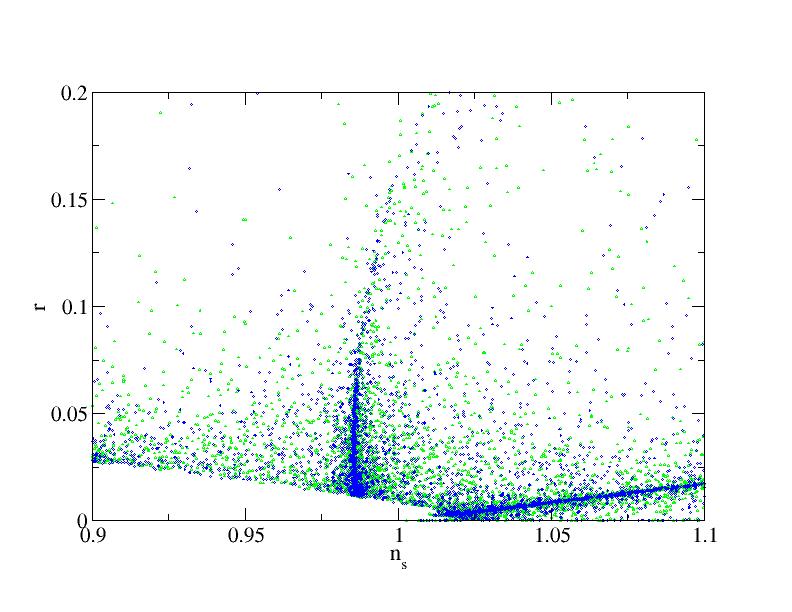}
\end{array}$
\caption{Monte Carlo results for conformal models (blue) and minimal models (green) satisfying $f_{NL} < 1$ (left panel) and $180 < f_{NL} < 200$ (right panel).  There is no detectable difference between the models in these ranges.}
\end{figure*}
\section{Results}\label{results}
We present the results of the Monte Carlo in
Fig. 1 for models that lie within approximate WMAP5 95\% CL (with running and tensors) \cite{Komatsu:2008hk}: $-0.1 < \alpha < 0.03$,
$0.9 <n_s < 1.1$, $-253 < f_{NL}^{\rm equil} < 151$~\footnote{A recent analysis finds $-435 < f_{NL}^{\rm equil} < 125$ \cite{Senatore:2009gt}.}, and $0 < r < 0.3$.  
We depict the $f_{NL}$-$\alpha$ plane and indicate the value
of $n_s$ with the color bar, for the conformal model (left panel) and the minimal model (right panel).  
There is a clear difference in the clustering of points between the
two models.  For the conformal model, we note the high density of points with red spectral index
($n_s < 1$) with negative running ($-0.1 \lesssim \alpha \lesssim 0$) and moderate degree of
non-Gaussianitiy ($1 \lesssim f_{NL} \lesssim 60$). 
If there are observable differences between the theories, it is in this intermediate
range that we expect the difference to be present, as the priors are chosen to ensure that the physics is the
same in the non-relativistic (low $f_{NL}$) and the relativistic (high $f_{NL}$) limits.
The lowest limit in this range compatible with the sensitivity expected from Planck for equilateral type
non-Gaussianity  is $f^{\rm equil}_{NL} \gtrsim 10$ \cite{Babich:2004yc}.  We will demonstrate that the most significant 
differences in models can be seen in the approximate range  $f_{NL} \in (20,40)$.

In order to demonstrate the accuracy of our numerical treatment, we verify that the minimal and conformal models do indeed give the same physics in the DBI and slow
roll limits.   In the left panel of Fig. 2, we present models in the $n_s$-$r$ plane with a large degree of non-Gaussianity: $180 < f^{\rm equil}_{NL} < 200$.  Conformal models are colored blue and minimal models colored green.  The two distributions of points overlap, indicating that the models are observationally degenerate in this range.  In the opposite limit, $f^{\rm equil}_{NL} < 1$, we again
see agreement, in the right panel of Fig. 2.  However, when we focus attention on the intermediate region, $20 < f^{\rm equil}_{NL} < 40$, in Fig. 3, we discover the surprising result
that the two models yield distinct observational predictions.
\begin{figure}
\centerline{\includegraphics[width=3.75in]{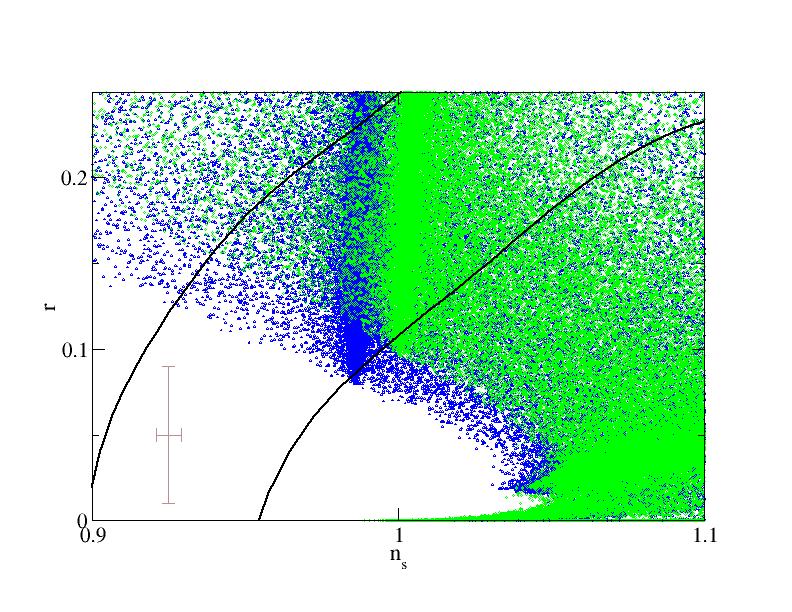}}
\caption{Monte Carlo results for conformal models (blue) and minimal models (green)
satisfying $20 < f_{NL} < 40$.  
The contours denote WMAP5+SDSS 1- and 2-$\sigma$ marginalized constraints.
The gray error bar is that expected from Planck at 1-$\sigma$.}
\end{figure}
We find that there is an observable region that is populated by only one of the
two classes of models; the conformal models can accommodate a redder spectral index for a given $r$ and $f_{NL}$.  
Additionally, the width
of this strip is of the same order as the 1-$\sigma$ error expected from Planck (gray error bar), suggesting that
these models may indeed be resolved in practice.
As we now show, the breaking of observable degeneracy is because of the difference in the field value at $N=55$ between the minimal and conformal models.

To begin, consider the following expressions for the DBI parameters and sound speed,
\begin{eqnarray}
\label{params} 
\epsilon & = & -\frac{\dot{H}}{H^2}
  = \frac{(\gamma+g_{,X})X}{M_{\rm Pl}^2H^2}, \nonumber\\
 \eta & = & \frac{\dot{\epsilon}}{H\epsilon}
  = \frac{1}{H}
  \left[\frac{\dot{\gamma}+\dot{g}_{,X}}{\gamma+g_{,X}}
   + \frac{\dot{X}}{X} - \frac{2\dot{H}}{H}
	 \right] \nonumber\\
 & = & \frac{1}{H}
  \left[\frac{\dot{\gamma}+\dot{g}_{,X}}{\gamma+g_{,X}}
   + \frac{2\gamma\dot{\gamma}}{\gamma^2-1}
   - \frac{2\dot{\gamma}}{\gamma}
   - \frac{\dot{f}}{f}
  \right] +2\epsilon, \nonumber\\
 s & = & \frac{\dot{c}_s}{Hc_s}
  = \frac{1}{2H}
  \left[\frac{\dot{\gamma}+\dot{g}_{,X}}{\gamma+g_{,X}}
   - \frac{3\gamma^2\dot{\gamma}+\dot{g}_{,X}}
   {\gamma^3+g_{,X}}\right], \nonumber\\
 c_s^2 & = & \frac{\gamma+g_{,X}}{\gamma^3+g_{,X}},
\end{eqnarray}
where we have taken $\tilde{\gamma} \approx \gamma$.  It is now a simple matter to generalize the Lyth bound of \cite{Lyth:1996im} (see Appendix \ref{genlyth}).

Using the expression for the spectral index, Eq.
(\ref{spindex}), and making various approximations valid for the regions of interest (see Appendix~\ref{specin}) we obtain
\begin{equation}
 n_s-1 \simeq -\frac{r\sqrt{3f_{NL}}}{4}
  + \frac{\sqrt{2r}}{\chi/M_{\rm Pl}}\,, \label{eqn:approxrelation}
\end{equation}
where $\chi$ is a general scalar field denoting either $\theta$ (minimal field) or $\Phi$ (conformal field).
\begin{figure}
\centerline{\includegraphics[width=3.75in]{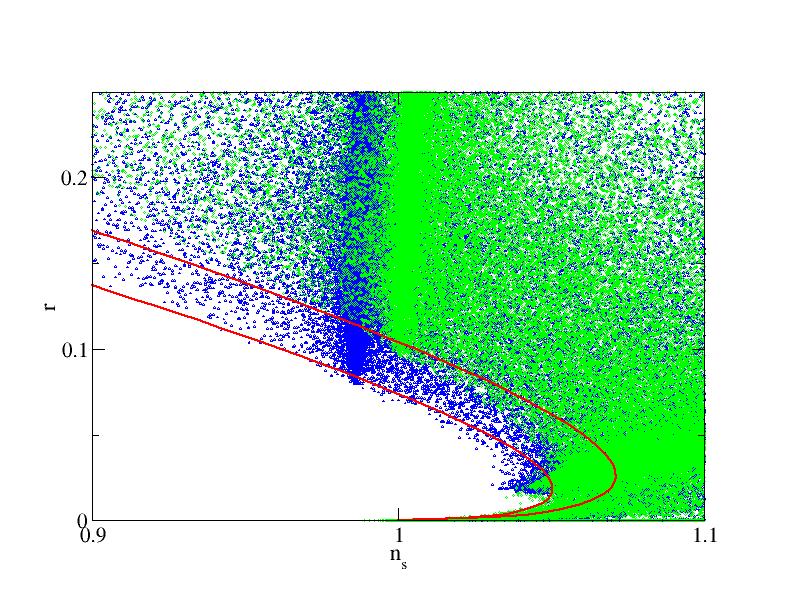}}
\caption{Same models as in Fig. 3, with theoretical curves obtained from Eq. (\ref{eqn:approxrelation}). The upper
red curve approximates the cut-off in minimal models, and the lower curve for conformal models.}
\end{figure}
\begin{figure}
\centerline{\includegraphics[width=3.75in]{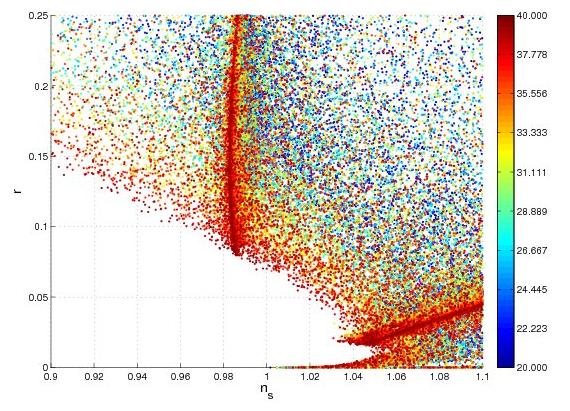}}
\caption{Same conformal models as in Fig. 3, with the value of $f_{NL}$ color-coded.}
\end{figure}
This result indicates that the difference seen in Fig. 3 is a result of the difference
between  $\theta_{55,{\rm max}}$ and
$\Phi_{55,{\rm max}}$, the maximum field values corresponding to $N=55$.  While this is only a rough approximation, it provides 
a reasonably accurate understanding of Fig. 3.  
By choosing $f_{NL} = 40$ and taking
the corresponding values of $\theta_{55,{\rm max}}$ and $\Phi_{55,{\rm max}}$ ($\Phi_{55,{\rm max}} >
\theta_{55,{\rm max}}$) in Eq. (\ref{eqn:approxrelation}), we obtain the curves approximating the cut-off of each
model in the $n_s$-$r$ plane (cf. Fig. 4). 
In order to justify our choice of constant $f_{NL}$ in this expression, in Fig. 5 we present
the same conformal models as in Fig. 3, but with the value of $f_{NL}$ color
coded.  Note that models with the same $f_{NL}$ tend to form bands in the
$n_s$-$r$ plane, increasing in magnitude towards small $n_s$.  These bands are
of the same form as the theoretical curves found in Fig. 4.   

It remains to understand why $\Phi_{55,{\rm max}} > \theta_{55,{\rm max}}$ in general.  
In the region under consideration, along the red curves in Fig.~4, we derive
the approximate relation 
\begin{equation}
fV \simeq  \frac{24}{r} \,, \label{eqn:fV}
\end{equation}
in Appendix~\ref{chi55}.
This expression is valid for both minimal and conformal models. 
We compare a minimal model with priors ($f(\theta)$, $V(\theta)$) and a
conformal model with ($f(\Phi)$, $V(\Phi_c)$). The approximate relation
Eq. (\ref{eqn:fV}) implies that 
%
\begin{equation}
 f(\Phi)V(\Phi_c) \simeq  f(\theta)V(\theta). 
\end{equation}
Noting that $V$ is an increasing function, while $fV$ is a decreasing 
function, $\Phi_c>\Phi$ implies that $\Phi$ tends to be larger than
$\theta$. 

As an illustration, let us consider the case where
%
\begin{equation}
 f(\theta) = \lambda\theta^{-4}, \quad 
  V(\theta) = V_0 + V_4\theta^4. 
\end{equation}
In this case we have
%
\begin{equation}
 \theta^{-4} \simeq \Phi^{-4}
  + \frac{V_4}{V_0}\left[\left(\frac{\Phi_c}{\Phi}\right)^4-1\right],
\end{equation}
and thus $\Phi>\theta$. Therefore, $\Phi$ tends to be larger than
$\theta$. Combining this with the result Eq. (\ref{eqn:approxrelation}), we
can say that, for $f_{NL}=40$, the minimum value of $n_s$ for a given
$r$ tends to be smaller for conformal models than for minimal models. 
\subsection{Negligible $r$ Regime}
\begin{figure}
\centerline{\includegraphics[width=3.75in]{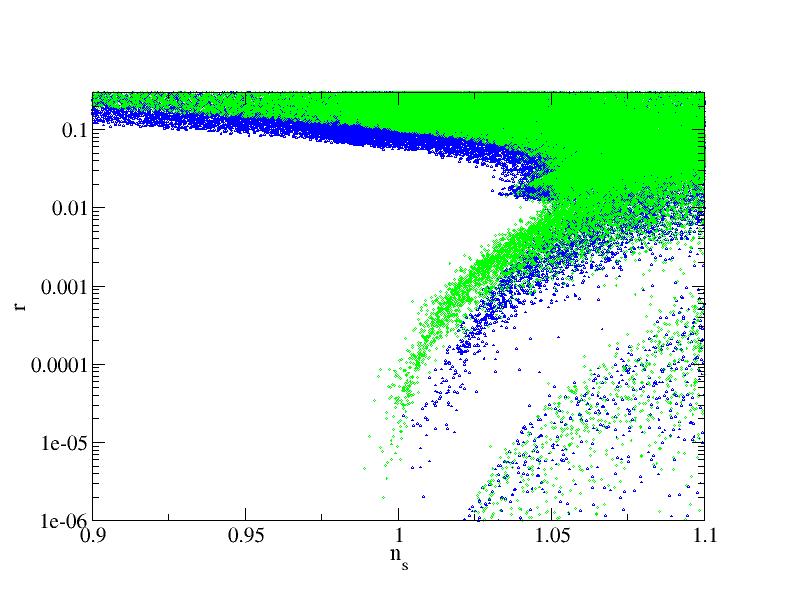}}
\caption{Close up of the negligible $r$ portion of the parameter space.}
\end{figure}
Comparing Fig.~3 and Fig.~5, our attention is drawn to another region
of potential interest near the Harrison-Zeldovich, $n_s =1$, $r=0$ point. Fig. 6
is a zoom of this region. To investigate this region of the $n_s$-$r$ plane we
are required to improve on our estimates of Appendix~\ref{specin}. We provide
a refined analysis to  $ {\mathcal O}(\gamma^{-1})$ in Appendix~\ref{ordergam}.
\subsubsection{Non-Minimally Coupled Models}
We begin with a discussion of the non-minimally coupled (blue) solutions in Fig. 6.
We note that there appear to be no blue solutions with a red spectral tilt and 
negligible tensor contribution $r\leq.06$.
This is easily understood from equation (\ref{eqn:nsconformal-betterformula}).
The last term in (\ref{eqn:nsconformal-betterformula}) is positive since $s$ is 
negative. This implies that the non-minimal models have blue spectra ($n_s>1$)
in the limit $r\to 0$.  Hence, any detection of $f_{NL}$ in the range $[20,40]$ together
with a measure of red spectral index and no detection of $r<.06$ would effectively
rule out all non-minimal DBI solutions (within our prior selection).

\subsubsection{Minimally Coupled Models}
We now turn our attention to the minimally coupled DBI models (green solutions).
This brings us to a discussion of the large white
region in Fig. 6 and the apparent absence of solutions with red spectral index and with low tensor signal $r\leq.06$.
As is evident from our large $f_{NL}$ analysis in Fig. 2, allowing models with significantly larger $f_{NL}  (> 40$) would gradually fill in this empty region.
The only successful red spectrum solutions are a small number of 
minimally coupled models located near the middle of the Figure. From (\ref{smrmm})
with $r \approx 0$ we find
\be
n_s - 1  \simeq  \frac{2s}{3f_{NL}}
\,,
\ee
where, in this region we have $s<0$ and $|s|={\mathcal O}(1)$ and
$\theta/M_{\rm Pl}={\mathcal O}(1)$. Hence, the reddest the solutions can be
are those with $f_{NL} = 20$, giving $n_s \simeq .97$. 
The spectrum
along the boundary of the populated region turns from blue to red at
$r=r_*$, where 
%
\begin{eqnarray}
 r_* \simeq \frac{1}{2}\left(\frac{2|s| \theta/M_{\rm Pl}}{3f_{NL}}\right)^2
  \simeq 10^{-4}. 
\end{eqnarray}
Hence, any detection of $f_{NL}$ in the range $[20,40]$ together
with a measure of red spectral index below $n_s \simeq .97$ and no detection of $r<.06$ would effectively
rule out all minimally coupled DBI solutions (within our prior selection).

Solutions in the bottom right corner typically have $\chi/M_{\rm Pl}={\mathcal O}(10^{-2})$,
and therefore have a blue spectral tilt since we can no longer ignore the second term in Eqs. (\ref{smrnm}), (\ref{smrmm}).
\section{Conclusions}\label{conclude}
In this paper we studied the effects of a non-minimally coupled inflaton to the 
DBI action. We assumed an AdS form for the warp factor and studied a wide range of potentials and parameter values using a Monte-Carlo analysis. Our aim was to determine the prospects for observationally distinguishing non-minimal coupling from standard minimally coupled DBI by adopting a phenomenological approach rather than to conform with the latest model building technology in string theory. Our first finding was that:
\begin{itemize}
\item
There is a strong degeneracy of observables in models with very large and models with undetectable levels of non-Gaussiantiy. 
\end{itemize}
This result is not particularly surprising given our analytic analysis of \S \ref{analytic}.
However, for an observationally interesting intermediate level of non-Gaussianity with $f_{NL} \in [20,40]$ we find that it is possible, in principle, to distinguish between minimally coupled and non-minimally coupled DBI models.~\footnote{Although our priors are carefully chosen to minimize the differences between the models, our conclusions are 
dependent upon these priors and the assumptions discussed in detail elsewhere in this paper.} \it Assuming a detection of intermediate levels of non-Gaussianity $f_{NL} \in [20,40]$ and the
value of the non-minimal coupling parameter $\xi = 1/6$ \rm we find:

\begin{itemize}
\item
For a given tensor-to-scalar ratio $r$ and $f_{NL}$, non-minimal DBI models tend to have a redder scalar spectral index $n_s$ 
than minimal DBI models. In other words,
for a given scalar spectral index $n_s$ and tensor-to-scalar ratio $r$, $f_{NL}$ tends to be smaller for non-minimal DBI models than for minimal DBI models.
\end{itemize}
Focusing our attention to the small tensor signal range we find:
\begin{itemize}
\item
Any detection of a red spectral index $n_s$ below .97 combined with no detection of a tensor signal at Planck sensitivity effectively would rule out
both non-minimal and minimal DBI inflation models.
\item
Any detection of a red spectrum and no detection of a tensor signal at Planck sensitivity would rule out non-minimally coupled DBI models.
\end{itemize}

Finally, the goal of this paper was to initiate the study of observable effects of gravitational couplings in theories with non-standard kinetic terms. While we have
focused on the particular example of conformal coupling and DBI kinetic term, our methods are applicable to more general systems represented by the 
action (\ref{act}). We leave further study of such models for future research.
\newpage
\acknowledgments
It is a pleasure to thank Ruth Gregory, Simeon Hellerman and Ivonne Zavala
for helpful conversations. This work is supported in part by the World Premier International Research 
Center Initiative (WPI Initiative), MEXT, Japan. DE is also supported by a Grant-in-Aid for Scientific Research 
(21740167) from the Japan Society for Promotion of Science (JSPS) and by funds from the Arizona State University Foundation.
SM is also supported by a Grant-in-Aid for Young Scientists (B) No. 17740134, by JSPS through a Grant-in-
Aid for Creative Scientific Research No. 19GS0219, and by the Mitsubishi Foundation.
%
%
\appendix
\section{Conformal Transformation}\label{ctrans}
%
Under the transformation
\begin{equation}
\tilde{g}_{\mu \nu} = \Omega^2 g_{\mu \nu},
\end{equation}
with $\Omega^2 =1-\xi\frac{\varphi^2}{M_{\rm Pl}^2}$,
the Ricci scalar transforms as
\begin{equation}
\tilde{R} = \Omega^{-2}\left[R + 3\left(\frac{\Omega'}{\Omega}\right)^2g^{\mu \nu}\partial_\mu \varphi \partial_\nu \varphi\right],
\end{equation}
where we have omitted a surface term.  From Eq. (\ref{ct}), the kinetic term transforms as
\begin{equation}
-\frac{1}{2}\sqrt{-g}g^{\mu \nu}\partial_\mu \varphi \partial_\nu \varphi =
-\frac{1}{2}\sqrt{-\tilde{g}}\Omega^{-2}\tilde{g}^{\mu \nu}\partial_\mu \varphi \partial_\nu \varphi.
\end{equation}
The conformal transformation thus results in a modified kinetic term,
\begin{eqnarray}
-\frac{1}{2}\sqrt{-\tilde{g}}\mathcal{F}\tilde{g}^{\mu \nu}\partial_\mu \varphi \partial_\nu \varphi,\\
\mathcal{F}= 6\left(\frac{\Omega'}{\Omega}\right)^2 + \Omega^{-2}.
\end{eqnarray}
The kinetic term can be made canonical by introducing a new field, $\sigma$, satisfying
\begin{equation}
\partial_\mu \sigma = -\sqrt{\mathcal{F}} \partial_\mu \varphi.
\end{equation}
In terms of this new field, we obtain the Einstein frame action,
\begin{equation}
S = \int d^4x \sqrt{-\tilde{g}}\left\{\frac{M_{\rm Pl}^2}{2}\tilde{R} - \frac{1}{2}\tilde{g}^{\mu
\nu}\partial_\mu \sigma \partial_\nu \sigma - \tilde{V}(\sigma)\right\},
\end{equation}
where $\tilde{V}(\sigma) = V(\varphi)\Omega^{-4}(\varphi)$.
\section{Spectral Index}\label{specin}
From the expression for the spectral index, Eq.
(\ref{spindex}), we obtain
\begin{eqnarray}
 -(n_s-1)-4\epsilon & = & (\eta-2\epsilon) + s \nonumber\\
 & = & 
  \left[-\frac{3g_{,X}(\gamma^2-1)\gamma}{2(\gamma+g_{,X})(\gamma^3+g_{,X})}\right.\nonumber \\
 & +&\left.\frac{2}{\gamma^2-1}\right]\frac{\dot{\gamma}}{H\gamma}
 + \frac{3\gamma^3-\gamma+2g_{,X}}{2(\gamma+g_{,X})(\gamma^3+g_{,X})}\nonumber \\
 &\times&\frac{\dot{g}_{,X}}{H}
 - \frac{\dot{f}}{Hf} \nonumber\\
 & = & -\frac{3g_{,X}\dot{\gamma}}{2H\gamma^2}\times A
  - \frac{\dot{f}}{Hf}\times B,
\end{eqnarray}
where 
%
\begin{eqnarray}\label{eqn:AandB}
 A & = & \frac{1-\gamma^{-2}}{(1+\gamma^{-1}g_{,X})(1+\gamma^{-3}g_{,X})}
  - \frac{4}{3g_{,X}\gamma}\cdot\frac{1}{1-\gamma^{-2}}, \nonumber\\
 B & = & 1 -
  \frac{3g_{,X}}{2\gamma}\cdot\frac{g_{X}'\chi}{g_{,X}}\cdot\frac{f}{f'\chi}
  \cdot\frac{1-\frac{1}{3}\gamma^{-2}+\frac{2}{3}\gamma^{-3}g_{,X}}
  {(1+\gamma^{-1}g_{,X})(1+\gamma^{-3}g_{,X})},\nonumber \\  
\end{eqnarray}
where $\chi$ is a general scalar field denoting either $\theta$ (minimal field) or $\Phi$ (conformal field). Because of the numerical facts
%
\begin{equation}
 \gamma=\mathcal{O}(10), \,\,  g_{,X}=\mathcal{O}(1), \,\,
 \frac{g_{,X}'\chi}{g_{,X}}=\mathcal{O}(1), \,\, \frac{f'\chi}{f}=\mathcal{O}(1)
\end{equation}
we can safely set
%
\begin{equation}
 A \simeq 1, \quad B \simeq 1, \quad
  \frac{\dot{\gamma}}{H\gamma^2} \simeq - c_s s.
\end{equation}
Thus, 
%
\begin{equation}
 -(n_s-1)-4\epsilon \simeq 
  \frac{3}{2}g_{,X}c_ss - \frac{\dot{f}}{Hf}
  = \frac{3}{2}g_{,X}c_ss + \frac{4\dot{\chi}}{H\chi},
\end{equation}
where we have used the ansatz $f\propto \chi^{-4}$. As another
numerical fact, we know that the first term in the final expression is
small ($g_{,X}=\mathcal{O}(1)$, $c_s=\mathcal{O}(0.1)$, $s=\mathcal{O}(0.1)$). In order to rewrite the
second term in terms of observables (and $\chi/M_{\rm Pl}$), we can use the
relation 
%
\begin{equation}
 \left(\frac{\dot{\chi}}{H\chi}\right)^2 = \frac{2X}{H^2\chi^2}
  = \frac{2\epsilon}{\gamma+g_{,X}}\cdot\left(\frac{M_{\rm Pl}}{\chi}\right)^2 
  \simeq \frac{2c_s\epsilon}{(\chi/M_{\rm Pl})^2}. 
\end{equation}
Therefore, for $\chi>0$ and $\dot{\chi}<0$, 
%
\begin{equation}
 n_s-1 \simeq -4\epsilon + \frac{\sqrt{32c_s\epsilon}}{\chi/M_{\rm Pl}}. 
\end{equation}
From the expression 
%
\begin{equation}
 r = 16c_s\epsilon,
\end{equation}
we obtain
%
\begin{equation}
 n_s-1 \simeq - \frac{r}{4c_s}  + \frac{\sqrt{2r}}{\chi/M_{\rm Pl}}. 
\end{equation}
Finally, by using the formula
%
\begin{equation}
 f_{NL} \simeq \frac{1}{3c_s^2}, 
\end{equation}
we find
%
\begin{equation}
 n_s-1 \simeq -\frac{r\sqrt{3f_{NL}}}{4}
  + \frac{\sqrt{2r}}{\chi/M_{\rm Pl}}.
\end{equation}
\section{Generalized Lyth Bound}\label{genlyth}
The generalized Lyth bound is derived from (\ref{eqn:r_cs_epsilon}) and
(\ref{params}):
%
\begin{equation}
 r = \frac{8}{M_{\rm Pl}^2}
  \sqrt{\frac{(1+\gamma^{-1}g_{,X})^3}{1+\gamma^{-3}g_{,X}}}
  \left(\frac{1}{H}\frac{d\chi}{dt}\right)^2.
\end{equation}
Thus, we obtain
%
\begin{eqnarray}
 \chi(N=55) - \chi(N=0) && \nonumber \\
=  \frac{M_{\rm Pl}}{2\sqrt{2}}\int_{\chi_0}^{\chi_{55}}
  \left[\frac{1+\gamma^{-3}g_{,X}}{(1+\gamma^{-1}g_{,X})^3}\right]^{1/4}
  \sqrt{r}dN  \nonumber \\
  \simeq 
  \frac{M_{\rm Pl}}{2\sqrt{2}}\int_{\chi_0}^{\chi_{55}}
  \left(1-\frac{3}{4}\gamma^{-1}g_{,X}\right)\sqrt{r}dN,
\end{eqnarray}
where $dN=-Hdt$.
\section{Derivation of Relation (\ref{eqn:fV})}\label{chi55}
By using Eqs. (\ref{params}), (\ref{eom}) and
(\ref{g}), we find
%
\begin{eqnarray}
 \frac{3(\gamma+g_{,X})}{\epsilon} &=& \frac{3M_{\rm Pl}^2H^2}{X}
  = \frac{1}{fX}\left[\gamma-1+fV\right]+g_{,X} \nonumber \\
  &=& \frac{2\gamma^2}{\gamma^2-1}\left[\gamma-1+fV\right]+g_{,X}.
\end{eqnarray}
Thus, 
%
\begin{equation}
\label{FV}
 fV = \frac{\gamma^2-1}{2\gamma^2}
  \left[ \frac{3(\gamma+g_{,X})}{\epsilon}-g_{,X}\right]
  - \gamma + 1. 
\end{equation}
From
%
\begin{equation}
 \gamma = \frac{1}{c_s}\sqrt{\frac{1+\gamma^{-1}g_{,X}}{1+\gamma^{-3}g_{,X}}}, 
  \quad
  \epsilon = \frac{r}{16c_s}, 
\end{equation}
we can rewrite Eq. (\ref{FV}) as
%
\begin{eqnarray}
 fV &=& 
  \left[ 1 - c_s^2\frac{1+\gamma^{-3}g_{,X}}{1+\gamma^{-1}g_{,X}}\right]\cdot
  \left[\frac{24}{r}\sqrt{\frac{(1+\gamma^{-1}g_{,X})^3}{1+\gamma^{-3}g_{,X}}}
   -\frac{g_{,X}}{2} \right] \nonumber \\
  &-&\frac{1}{c_s}\sqrt{\frac{1+\gamma^{-1}g_{,X}}{1+\gamma^{-3}g_{,X}}} + 1. 
\end{eqnarray}
This result applies to conformal models. On the other hand, for minimal models we
have 
%
\begin{equation}
 fV = (1 - c_s^2)\frac{24}{r}-\frac{1}{c_s} + 1
  = \left(1-\frac{1}{3f_{NL}}\right)\frac{24}{r}-\sqrt{3f_{NL}} + 1. 
\end{equation}
This means that, for minimal models, the microscopic quantity $fV$ is
expressed in terms of observables $r$ and $f_{NL}$. 

Now, for $f_{NL}=40$, the numerical facts 
%
\begin{equation}
 c_s \simeq \gamma^{-1} \simeq 0.1, \quad g_{,X} = \mathcal{O}(1)
\end{equation}
lead us to a rough approximation
%
\begin{equation}
fV \simeq  \frac{24}{r} \,,
\end{equation}
for small $r$.
\section{Corrections of Order $ {\mathcal O}(\gamma^{-1})$}\label{ordergam}
\subsection{Non-Minimally Coupled Models}
%
In order to improve accuracy of the formula (\ref{eqn:approxrelation})
we can include corrections first-order in $\gamma^{-1}$. The result is  
%
\begin{eqnarray}
 n_s &-& 1 \simeq -\frac{r\sqrt{3f_{NL}}}{4}
  - \frac{f'\Phi}{4f}\cdot\frac{\sqrt{2r}}{\Phi/M_{\rm Pl}} \times  \nonumber \\
&&  \left\{1 + \frac{3g_{,X}}{4\sqrt{3f_{NL}}}\cdot
   \left[\frac{g'\Phi}{2g_{,X}}\cdot\left(-\frac{4f}{f'\Phi}\right)-1\right]\right\} \nonumber \\ 
 && - \frac{3g_{,X}s}{2\sqrt{3f_{NL}}}. 
\end{eqnarray}
Here, we have used
%
\begin{eqnarray}
 B & \simeq & 1 - \frac{3}{2}\gamma^{-1}g_{,X}
  \cdot\frac{g_{,X}'\Phi}{g_{,X}}\cdot\frac{f}{f'\Phi}, \nonumber\\
 c_s & \simeq & \gamma^{-1}\left(1+\frac{1}{2}\gamma^{-1}g_{,X}\right),
  \nonumber\\
 \left(\frac{\dot{\Phi}}{H \Phi}\right)^2 & \simeq & 
  \frac{2c_s\epsilon}{(\Phi/M_{\rm Pl})^2}\times
  \left(1 - \frac{3}{2}\gamma^{-1}g_{,X}\right). 
\end{eqnarray}
By using $f\propto \Phi^{-4}$ and $g_{,X}=\tan^2(\Phi/\sqrt{6}M_{\rm Pl})$, we
obtain 
%
\begin{eqnarray}\label{smrnm}
 n_s &-& 1  \simeq  -\frac{r\sqrt{3f_{NL}}}{4} 
  +  \frac{\sqrt{2r}}{\Phi/M_{\rm Pl}} \times \nonumber \\
 && \left\{1 + \frac{3g_{,X}}{4\sqrt{3f_{NL}}}
   \left[
    \frac{(2 \Phi/\sqrt{6}M_{\rm Pl})}{\sin(2 \Phi/\sqrt{6}M_{\rm Pl})}
    -1\right]\right\}  \nonumber \\
  && -  \frac{3g_{,X}s}{2\sqrt{3f_{NL}}}. 
  \label{eqn:nsconformal-betterformula}
\end{eqnarray}
%
\subsection{Minimally Coupled Models}
The corresponding relation for minimially coupled models can be obtained
by setting $g_{,X}=0$. However, we have to be careful about the last term in
$A$ shown in (\ref{eqn:AandB}). Then we obtain
%
\begin{eqnarray}
 -(n_s-1)-4\epsilon = 
  \frac{2\dot{\gamma}}{H\gamma^3}\cdot\frac{1}{1-\gamma^{-2}}
  - \frac{\dot{f}}{Hf}. 
\end{eqnarray}
This leads to the following approximate relation
%
\begin{eqnarray}\label{smrmm}
 n_s-1 \simeq -\frac{r\sqrt{3f_{NL}}}{4}
  + \frac{\sqrt{2r}}{\theta/M_{\rm Pl}} + \frac{2s}{3f_{NL}}.
\end{eqnarray}
Interestingly enough, minimally coupled models have red spectra 
($n_s<1$) in the limit $r\to 0$ since $s$ is negative. 
Here, we have used numerical values $20<f_{NL}<40$, $|s|= {\mathcal O}(1)$ and
$\theta/M_{\rm Pl}= {\mathcal O}(1)$. 



\begin{thebibliography}{99}
\bibitem{Guth:1980zm}
  A.~H.~Guth,
  Phys.\ Rev.\  D {\bf 23}, 347 (1981).
  
\bibitem{Linde:1981mu}
  A.~D.~Linde,
  Phys.\ Lett.\  B {\bf 108}, 389 (1982).
 
\bibitem{Albrecht:1982wi}
  A.~Albrecht and P.~J.~Steinhardt,
  Phys.\ Rev.\ Lett.\  {\bf 48}, 1220 (1982).
  
  \bibitem{Komatsu:2008hk}
  E.~Komatsu {\it et al.}  [WMAP Collaboration],
  Astrophys.\ J.\ Suppl.\  {\bf 180}, 330 (2009)
  [arXiv:0803.0547 [astro-ph]].
  
\bibitem{Brandenberger:2007qi}
  R.~H.~Brandenberger,
  Lect.\ Notes Phys.\  {\bf 738}, 393 (2008)
  [arXiv:hep-th/0701111].
  
\bibitem{Dvali:1998pa}
  G.~R.~Dvali and S.~H.~H.~Tye,
  Phys.\ Lett.\  B {\bf 450}, 72 (1999)
  [arXiv:hep-ph/9812483].
  
\bibitem{Alexander:2001ks}
  S.~H.~S.~Alexander,
  Phys.\ Rev.\  D {\bf 65}, 023507 (2002)
  [arXiv:hep-th/0105032].
  
\bibitem{Dvali:2001fw}
  G.~R.~Dvali, Q.~Shafi and S.~Solganik,
  arXiv:hep-th/0105203.
  
\bibitem{Burgess:2001fx}
  C.~P.~Burgess, M.~Majumdar, D.~Nolte, F.~Quevedo, G.~Rajesh and R.~J.~Zhang,
  JHEP {\bf 0107}, 047 (2001)
  [arXiv:hep-th/0105204].

\bibitem{Brodie:2003qv}
  J.~H.~Brodie and D.~A.~Easson,
  JCAP {\bf 0312}, 004 (2003)
  [arXiv:hep-th/0301138].
  
\bibitem{Kachru:2003sx}
  S.~Kachru, R.~Kallosh, A.~Linde, J.~M.~Maldacena, L.~P.~McAllister and S.~P.~Trivedi,
  JCAP {\bf 0310}, 013 (2003)
  [arXiv:hep-th/0308055].
  
\bibitem{McAllister:2007bg}
  L.~McAllister and E.~Silverstein,
  Gen.\ Rel.\ Grav.\  {\bf 40}, 565 (2008)
  [arXiv:0710.2951 [hep-th]].
  
\bibitem{Baumann:2009ni}
  D.~Baumann and L.~McAllister,
  arXiv:0901.0265 [hep-th].
  
\bibitem{ArmendarizPicon:1999rj}
  C.~Armendariz-Picon, T.~Damour and V.~F.~Mukhanov,
  Phys.\ Lett.\  B {\bf 458}, 209 (1999)
  [arXiv:hep-th/9904075].
  
\bibitem{ArkaniHamed:2003uz}
  N.~Arkani-Hamed, P.~Creminelli, S.~Mukohyama and M.~Zaldarriaga,
  JCAP {\bf 0404}, 001 (2004)
  [arXiv:hep-th/0312100].

  
  \bibitem{Silverstein:2003hf}
  E.~Silverstein and D.~Tong,
  Phys.\ Rev.\  D {\bf 70}, 103505 (2004)
  [arXiv:hep-th/0310221].
  
  
\bibitem{Garriga:1999vw}
  J.~Garriga and V.~F.~Mukhanov,
  Phys.\ Lett.\  B {\bf 458}, 219 (1999).
 
\bibitem{Alishahiha:2004eh}
  M.~Alishahiha, E.~Silverstein and D.~Tong,
  Phys.\ Rev.\  D {\bf 70}, 123505 (2004)
  [arXiv:hep-th/0404084].
  
 \bibitem{Chen:2006nt}
  X.~Chen, M.~x.~Huang, S.~Kachru and G.~Shiu,
  JCAP {\bf 0701}, 002 (2007)
  [arXiv:hep-th/0605045].
  
\bibitem{Birrell:1982ix}
  N.~D.~Birrell and P.~C.~W.~Davies,
{\it  Cambridge, Uk: Univ. Pr. ( 1982) 340p}.
 
  
  

\bibitem{Seiberg:1999xz}
  N.~Seiberg and E.~Witten,
  JHEP {\bf 9904}, 017 (1999)
  [arXiv:hep-th/9903224].
  
\bibitem{Fotopoulos:2002wy}
  A.~Fotopoulos and A.~A.~Tseytlin,
  JHEP {\bf 0212}, 001 (2002)
  [arXiv:hep-th/0211101].
  
\bibitem{Burgess:2009ea}
  C.~P.~Burgess, H.~M.~Lee and M.~Trott,
  JHEP {\bf 0909}, 103 (2009)
  [arXiv:0902.4465 [hep-ph]].
  
\bibitem{KSgeom}
  I.~R.~Klebanov and M.~J.~Strassler,
  JHEP {\bf 0008}, 052 (2000)
  [arXiv:hep-th/0007191].
  
\bibitem{Chen:2006hs}
  X.~Chen, S.~Sarangi, S.~H.~Henry Tye and J.~Xu,
  JCAP {\bf 0611}, 015 (2006)
  [arXiv:hep-th/0608082].
  
\bibitem{Baumann:2006cd}
  D.~Baumann and L.~McAllister,
  Phys.\ Rev.\  D {\bf 75}, 123508 (2007)
  [arXiv:hep-th/0610285].
  
\bibitem{Peiris:2007gz}
  H.~V.~Peiris, D.~Baumann, B.~Friedman and A.~Cooray,
  Phys.\ Rev.\  D {\bf 76}, 103517 (2007)
  [arXiv:0706.1240 [astro-ph]].
  
\bibitem{Easson:2007dh}
  D.~A.~Easson, R.~Gregory, D.~F.~Mota, G.~Tasinato and I.~Zavala,
  JCAP {\bf 0802}, 010 (2008)
  [arXiv:0709.2666 [hep-th]].
  
\bibitem{Bean:2007eh}
  R.~Bean, X.~Chen, H.~Peiris and J.~Xu,
  Phys.\ Rev.\  D {\bf 77}, 023527 (2008)
  [arXiv:0710.1812 [hep-th]].

\bibitem{Leblond:2008gg}
  L.~Leblond and S.~Shandera,
  JCAP {\bf 0808}, 007 (2008)
  [arXiv:0802.2290 [hep-th]].
  
\bibitem{Becker:2007ui}
  M.~Becker, L.~Leblond and S.~E.~Shandera,
  Phys.\ Rev.\  D {\bf 76}, 123516 (2007)
  [arXiv:0709.1170 [hep-th]].
     
\bibitem{Silverstein:2008sg}
  E.~Silverstein and A.~Westphal,
  Phys.\ Rev.\  D {\bf 78}, 106003 (2008)
  [arXiv:0803.3085 [hep-th]].
  
\bibitem{McAllister:2008hb}
  L.~McAllister, E.~Silverstein and A.~Westphal,
  arXiv:0808.0706 [hep-th].

\bibitem{Avgoustidis:2008zu}
  A.~Avgoustidis and I.~Zavala,
  JCAP {\bf 0901}, 045 (2009)
  [arXiv:0810.5001 [hep-th]].
    
\bibitem{Starobinsky81}
  A.~A.~Starobinsky,
Pis'ma \ Astron.\ Zh. {\bf 7}, 67 (1981).
  
\bibitem{Futamase:1989hb}
  T.~Futamase, T.~Rothman and R.~Matzner,
  Phys.\ Rev.\  D {\bf 39}, 405 (1989).
  
\bibitem{Makino:1991sg}
  N.~Makino and M.~Sasaki,
  Prog.\ Theor.\ Phys.\  {\bf 86}, 103 (1991).
  
\bibitem{Creminelli:2003iq}
  P.~Creminelli,
  JCAP {\bf 0310}, 003 (2003)
  [arXiv:astro-ph/0306122].

\bibitem{Bean:2007hc}
  R.~Bean, S.~E.~Shandera, S.~H.~Henry Tye and J.~Xu,
  JCAP {\bf 0705}, 004 (2007)
  [arXiv:hep-th/0702107].

\bibitem{Maldacena:2002vr}
  J.~M.~Maldacena,
  JHEP {\bf 0305}, 013 (2003)
  [arXiv:astro-ph/0210603].
  
\bibitem{Gangui:1993tt}
  A.~Gangui, F.~Lucchin, S.~Matarrese and S.~Mollerach,
  Astrophys.\ J.\  {\bf 430}, 447 (1994)
  [arXiv:astro-ph/9312033].

\bibitem{Komatsu:2001rj}
  E.~Komatsu and D.~N.~Spergel,
  Phys.\ Rev.\  D {\bf 63}, 063002 (2001)
  [arXiv:astro-ph/0005036].

\bibitem{Babich:2004gb}
  D.~Babich, P.~Creminelli and M.~Zaldarriaga,
  JCAP {\bf 0408}, 009 (2004)
  [arXiv:astro-ph/0405356].
  
\bibitem{Fergusson:2008ra}
  J.~R.~Fergusson and E.~P.~S.~Shellard,
  arXiv:0812.3413 [astro-ph].
  
\bibitem{Easson:2009kk}
  D.~A.~Easson and R.~Gregory,
  Phys.\ Rev.\  D {\bf 80}, 083518 (2009)
  [arXiv:0902.1798 [hep-th]].
 
\bibitem{Easther:2002rw}
  R.~Easther and W.~H.~Kinney,
  Phys.\ Rev.\  D {\bf 67}, 043511 (2003)
  [arXiv:astro-ph/0210345].
  
\bibitem{Powell:2007gu}
  B.~A.~Powell and W.~H.~Kinney,
  JCAP {\bf 0708}, 006 (2007)
  [arXiv:0706.1982 [astro-ph]].
  
\bibitem{Lorenz:2008je}
  L.~Lorenz, J.~Martin and C.~Ringeval,
  Phys.\ Rev.\  D {\bf 78}, 063543 (2008)
  [arXiv:0807.2414 [astro-ph]].
  
\bibitem{Lorenz:2008et}
  L.~Lorenz, J.~Martin and C.~Ringeval,
  Phys.\ Rev.\  D {\bf 78}, 083513 (2008)
  [arXiv:0807.3037 [astro-ph]].
  
  
\bibitem{Agarwal:2008ah}
 N.~Agarwal and R.~Bean,
 Phys.\ Rev.\  D {\bf 79}, 023503 (2009)
 [arXiv:0809.2798 [astro-ph]].
    
  \bibitem{Powell:2008bi}
  B.~A.~Powell, K.~Tzirakis and W.~H.~Kinney,
  JCAP {\bf 0904}, 019 (2009)
  [arXiv:0812.1797 [astro-ph]].
  
\bibitem{Senatore:2009gt}
  L.~Senatore, K.~M.~Smith and M.~Zaldarriaga,
  arXiv:0905.3746 [astro-ph.CO].

\bibitem{Lyth:1996im}
  D.~H.~Lyth,
  Phys.\ Rev.\ Lett.\  {\bf 78}, 1861 (1997)
  [arXiv:hep-ph/9606387].
  
  
\bibitem{Babich:2004yc}
  D.~Babich and M.~Zaldarriaga,
  Phys.\ Rev.\  D {\bf 70}, 083005 (2004)
  [arXiv:astro-ph/0408455].
  

 

\end{thebibliography}
\end{document}